\documentclass[12pt]{iopart}

\usepackage[dvipsnames]{xcolor}
\usepackage{hyperref}
\hypersetup{
    colorlinks,
    linkcolor=NavyBlue,
    citecolor=BrickRed,
    urlcolor=MidnightBlue
}
\usepackage{iopams}
\usepackage{graphicx} 
\usepackage{amsmath}
\usepackage{dcolumn} 
\usepackage{bm} 
\usepackage{hyperref}
\usepackage{nth}
\usepackage[nointegrals]{wasysym}
\usepackage{physics}
\usepackage{ulem}
\usepackage{caption}
\usepackage{svg}
\usepackage{siunitx}

\newcommand{\vect}[1]{\boldsymbol{#1}}

\newcommand{\NA}{\mathrm{NA}}
\newcommand{\exc}{\mathrm{exc}}
\newcommand{\emi}{\mathrm{em}}

\newcommand{\FT}{\mathcal{F}}

\newcommand{\xs}{\vect x_s}
\newcommand{\xd}{\vect x_d}

\DeclareMathOperator*{\argmin}{arg\,min}
\DeclareMathOperator*{\argmax}{arg\,max}

\begin{document}

\title[Image Scanning Microscopy Reconstruction by Autocorrelation Inversion]
      {Image Scanning Microscopy Reconstruction by Autocorrelation Inversion}

\author{Daniele Ancora$^{1,2,\dagger,*}$,
Alessandro Zunino$^{3,*}$,
Giuseppe Vicidomini$^3$,
Alvaro H. Crevenna$^1$}
\address{$^1$European Molecular Biological Laboratory, Epigenetics and Neurobiology unit, Rome - Italy}
\address{$^2$Consiglio Nazionale delle Ricerce, Institute of Nanotechnology, Rome - Italy}
\address{$^3$Italian Institute of Technology, Genoa - Italy}
\address{*\textit{these authors equally contributed to the manuscript}.}
\ead{$^\dagger$daniele.ancora@embl.it}

\vspace{10pt}

\begin{abstract}
Confocal laser scanning microscopy (CLSM) stands out as one of the most widely used microscopy techniques thanks to its three-dimensional imaging capability and its sub-diffraction spatial resolution, achieved through the closure of a pinhole in front of a single-element detector. However, the pinhole also rejects useful photons, and beating the diffraction limit comes at the price of irremediably compromising the signal-to-noise ratio (SNR) of the data. Image scanning microscopy (ISM) emerged as the rational evolution of CLSM, exploiting a small array detector in place of the pinhole and the single-element detector. Each sensitive element is small enough to achieve sub-diffraction resolution through the confocal effect, but the size of the whole detector is large enough to guarantee excellent collection efficiency and SNR. However, the raw data produced by an ISM setup consists of a 4D dataset, which can be seen as a set of confocal-like images. Thus, fusing the dataset into a single super-resolved image requires a dedicated reconstruction algorithm. Conventional methods are multi-image deconvolution, which requires prior knowledge of the system point spread functions (PSF), or adaptive pixel reassignment (APR), which is effective only on a limited range of experimental conditions. In this work, we describe and validate a novel concept for ISM image reconstruction based on autocorrelation inversion. We leverage unique properties of the autocorrelation to discard low-frequency components and maximize the resolution of the reconstructed image without any assumption on the image or any knowledge of the PSF. Our results push the quality of the ISM reconstruction beyond the level provided by APR and open new perspectives for multi-dimensional image processing.
\end{abstract}

\section{Introduction}

Microscopy is an indispensable tool in scientific research, offering a window into the microscopic world that shapes our understanding of biology, chemistry, and material science. Among the many microscopy techniques developed over the years, image scanning microscopy (ISM) stands out as a modern evolution of conventional confocal laser scanning microscopy (CLSM).
In traditional fluorescence CLSM, a focused excitation beam scans a sample while a single-element detector captures the resulting fluorescence light. A small pinhole sits in the image plane of the confocal microscope, blocking the out-of-focus light from reaching the detector and achieving optical sectioning \cite{Sheppard1977}. Theoretically, a perfectly closed pinhole also enhances lateral resolution up to a factor of two compared to the diffraction limit, achieving super-resolution \cite{Sheppard1982, Sheppard1988}. However, the closer the pinhole, the less light the detector collects, inevitably compromising the image's signal-to-noise ratio (SNR) and making CLSM an ineffective technique to achieve super-resolution.
Fast and small array detectors \cite{Huff2017} started a paradigm shift in laser-scanning microscopy, enabling effective implementations of ISM. In ISM, array detectors replace the conventional pinhole and single-element detector of CLSM, capturing the spatial distribution of fluorescence emissions from each excited spot. The detector size is designed to be large enough to allow for good collection efficiency, but each sensitive element is small enough to function as a pinhole and enable super-resolution \cite{Muller2010}. These conditions make ISM a simple and user-friendly technique that achieves practical super-resolution with excellent SNR.

Despite its maturity, ISM is still a dynamic and quickly evolving research field, thanks to the recent introduction of innovative single photon avalanche diode (SPAD) array detectors, which add further versatility to the microscopes. Indeed, SPAD arrays can detect single photons with tens of picosecond temporal precision \cite{Buttafava2020}, enabling fluoresce lifetime ISM (FLISM) \cite{Castello2019, Rossetta2022, Tortarolo2024}, quantum ISM (Q-ISM) \cite{Tenne2019}, and super-resolution optical fluctuation ISM (SOFISM) \cite{Sroda2020}. Furthermore, ISM can be synergically combined with other scanning techniques, such as stimulated emission depletion (STED) \cite{Tortarolo2022} and multi-photon excitation microscopy \cite{Koho2020}.

From a conceptual standpoint, the ISM system can be interpreted as equivalent to multiple confocal microscopes, acquiring images of the same sample from slightly different points of views. Thus, the ISM setup produces a four-dimensional dataset of parallel-scanned images, one for each detector element. In order to form an image with enhanced SNR and lateral resolution, a tailored reconstruction algorithm is needed for the correct assemblage of this parallelized detection.
Assuming that the images of the ISM dataset are identical but shifted and rescaled in intensity, the pixel reassignment (PR) method aligns and sums the scanned images by applying inverse shifts estimated using the geometry of the detector array \cite{Sheppard2013}. Such an approach is so simple and effective that it enables real-time processing of images using analogue reassignment, either mechanical \cite{Roth2013, DeLuca2013} or optical \cite{York2013}.
On the downside, reassigning using constant shift values lacks versatility and poorly adapts to challenging scenarios where emission and excitation wavelength tunability,  misalignments or optical aberrations are present. A more robust approach to ISM reconstruction is adaptive pixel reassignment (APR) \cite{Castello2015, Castello2019, Sheppard2020}, which uses an image registration method (e.g., phase correlation) to estimate the shifts directly from the raw data without prior knowledge of the microscope's point spread function (PSF) or other imaging conditions. The APR approach is simple, fast, and can easily adapt to different scenarios, but it stands on an approximate assumption. Indeed, depending on the experimental parameters -- such as the magnification of the microscope and wavelength of light -- the PSFs of the images generated by the peripheral elements of the array detector might be not only shifted but also distorted in shape.
A more rigorous approach to ISM reconstruction is multi-image deconvolution \cite{Ingaramo2014, Zunino2023}, which consists of inverting the ISM image formation process by exploiting the exact shape of the scanned images' PSFs. Such an approach typically provides better results than APR but requires the additional knowledge of the system's PSF, which must be experimentally measured in a dedicated experiment or carefully simulated using diffraction theory \cite{Caprile2022, Zunino2023b}.

Here, we propose a new approach to ISM image reconstruction, which does not require any prior information or assumptions about the PSFs of the microscope.
The method -- named ACO-ISM -- is based on the inversion of the average autocorrelation (ACO) of the raw ISM dataset and can successfully reconstruct an image with enhanced SNR and lateral resolution, even when experimental conditions fail to fulfil the assumptions behind APR.
The idea behind ACO-ISM stems from the general observation that the inversion of the averaged autocorrelation is an efficient tool to fuse a collection of images of the same subject observed with different PSFs.
As recently demonstrated in the context of light-sheet microscopy, combining multiple-angle volumetric images by inverting the average autocorrelation produces a higher resolution reconstruction than with aligned fusion \cite{ancora2021beyond}.
Indeed, averaging images in the autocorrelation space can preserve the high-frequency content by discarding the cross-correlation terms that would degrade the resolution.
In the context of ISM, the physics of image formation implies that each image is blurred by a different PSF.
Therefore, a reconstruction procedure based on the autocorrelation inversion would be beneficial, especially when the experimental conditions maximize the PSF diversity.
The goal of this work is to enhance the super-resolution effect of ISM by developing a reconstruction algorithm that constructs and inverts the sharpest autocorrelation from the raw ISM dataset.
Since a narrow autocorrelation uniquely corresponds to a larger spatial spectrum, successfully inverting the autocorrelation problem produces an image with better spatial resolution.
In practice, we calculate the average of the autocorrelation of the raw ISM images, which we invert using the Schultz-Snyder \cite{schulz1992image} fixed-point iteration.
Although we rely on the fact that the PSFs are different between the different images, we do not require their knowledge, and neither do we attempt to estimate them blindly.
Granting access to a new reconstruction space, our strategy pushes even further the capabilities of ISM and extends its range of applicability also to previously acquired datasets.

\begin{figure}
    \centering
    \includegraphics[width=1.\textwidth]{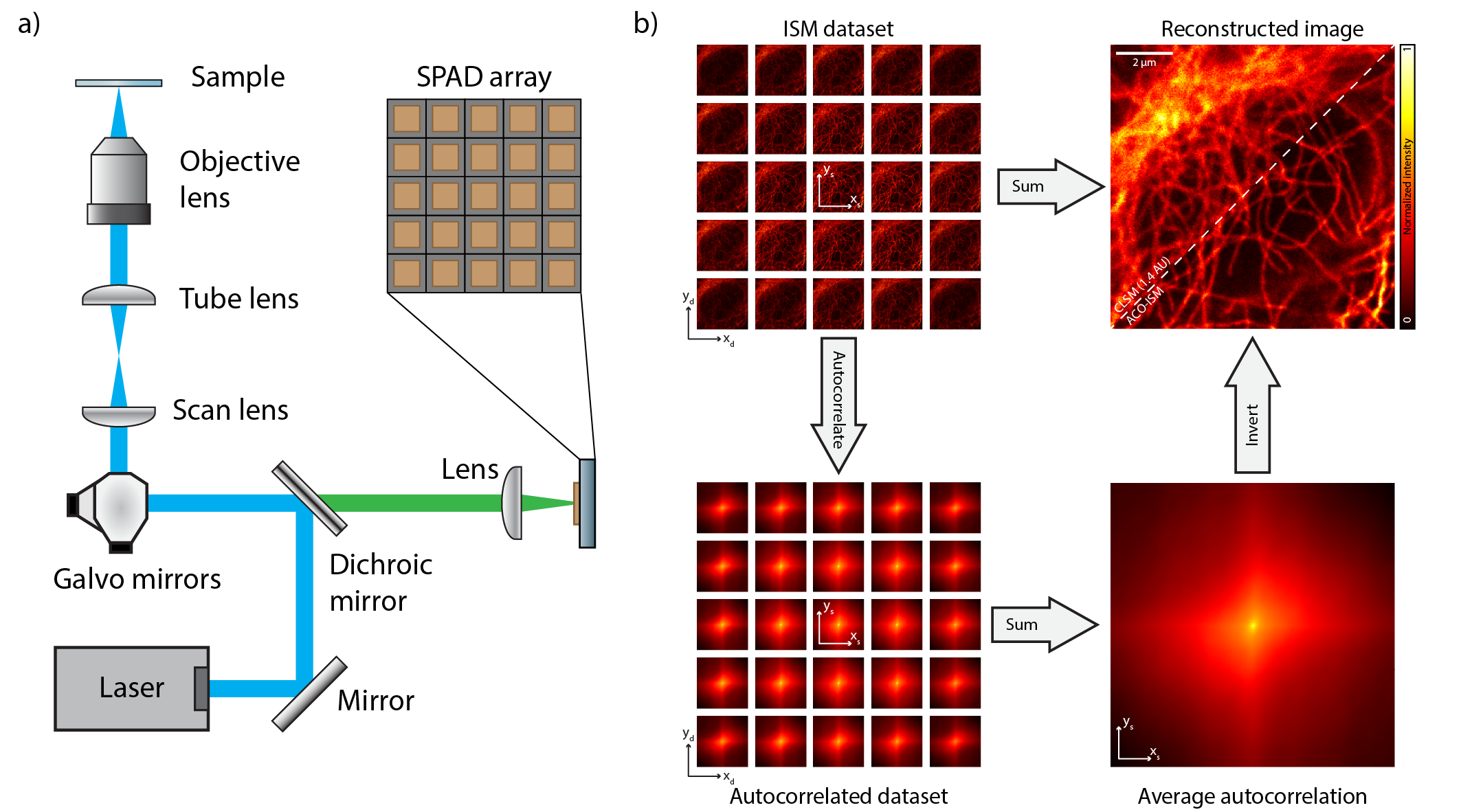}
    \caption{\textbf{Autocorrelation-based Image Scanning Microscopy} a) Sketch of an ISM setup, which consists of a laser scanning microscope equipped with an array detector.
    b) Image reconstruction procedure. Each image of the ISM dataset is autocorrelated and then summed. Finally, the average autocorrelation is inverted, to obtain a reconstruction with higher resolution and SNR compared to the diffraction-limited image.}
    \label{fig:opticalsetup}
\end{figure}

\section{Concept and implementation}
\subsection{Theoretical framework}
\label{sec:theory}

An ISM microscope investigates a sample by focusing a beam on the scan point $\xs = (x_s, y_s)$ of the sample plane and collecting the excited fluorescence light with a detector array (Fig. \ref{fig:opticalsetup}a). In this work, we used an asynchronous readout $5 \times 5$ SPAD array detector \cite{Buttafava2020}. Each sensitive element is located at the position $\xd = (x_d, y_d)$ on the detector/image plane and acts as a displaced pinhole. Thus, the ISM microscope returns a four-dimensional dataset $i(\xs|\xd)$ after a complete raster scan of the field-of-view, which can be seen as a collection of shifted confocal-like images. These latter describe the sample but are blurred by a different PSF. More in detail, the ISM image formation equation is
\begin{equation}
    i(\xs|\xd) = o(\xs) * h(\vect x_s | \vect x_d)
\end{equation}
where $*$ is the convolution operator with respect to the coordinate $\xs$, $o(\xs)$ is the object under investigation and $h(\vect x_s | \vect x_d)$ is the PSF corresponding to the detector element located at $\vect x_d$. More precisely, the set of PSFs is given by
\begin{equation}
    h(\vect x_s | \vect x_d) = h_{\exc}(-\vect x_s) \cdot \qty[ h_{\emi}(\vect x_s) * p(\vect x_s - \vect x_d) ]
\end{equation}
where $p$ is the function describing the geometry of the sensitive area of the detector, $h_{\exc}$ and $h_{\emi}$ are the excitation and emission PSF, respectively. Namely, the PSF of each detector element is given by the product of two diffraction-limited spots, hence the super-resolution effect. However, one PSF is shifted to the other by the quantity $\xd$. The smaller the shift, the more mutually similar the twenty-five PSFs are. The value of $\xd$ mainly depends on the magnification $M$ of the microscope. Since the detector has a finite size, a lower magnification corresponds to better photon collection efficiency but a more challenging reconstruction process, given the higher dissimilarity of the PSFs.

ISM processing is now routinely performed using the APR algorithm which assumes that the PSFs are structurally identical but shifted by a quantity $\vect\mu(\xd)$, named shift-vector \cite{Castello2015, Castello2019, Sheppard2020}. Thus, image reconstruction through the APR method consists of calculating and compensating the shifts and subsequently summing the aligned images. The result is:

\begin{equation}
    i_{\text{APR}}(\xs) = o(\xs) * \sum_{\xd} h(\xs + \vect \mu(\vect x_d) | \xd) = o(\xs) * h_{\text{APR}}(\xs)
\end{equation}
where $h_{\text{APR}}$ is sharper by a factor $\sqrt{2}$ with respect to the diffraction limit.
Failing to compensate for the shifts results in the loss of the super-resolution effect. Indeed, the effect of summing all the images of the dataset is:

\begin{equation}
    i_{\text{SUM}}(\xs) = o(\xs) * \sum_{\xd} h(\xs | \xd) = o(\xs) * h_{\text{SUM}}(\xs)
\end{equation}
where $h_{\text{SUM}}$ is the conventional diffraction-limited excitation PSF. Namely, the resulting image is the same obtained by a confocal microscope with an open pinhole.

\begin{figure}
    \centering
    \includegraphics[width=1.\textwidth]{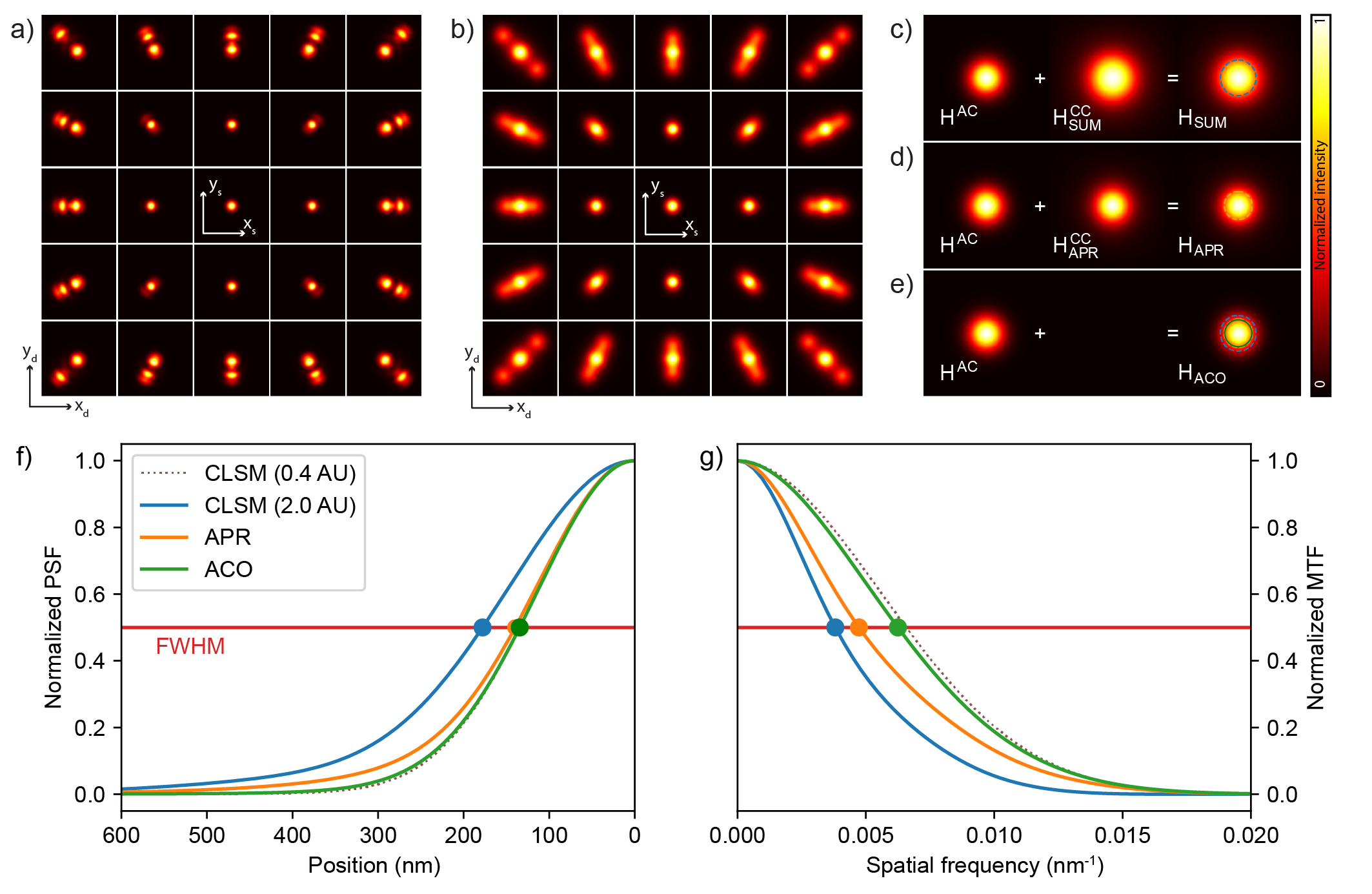}
    \caption{\textbf{Working principle of ISM reconstruction through autocorrelation inversion}. Panel a), ISM dataset of PSFs simulated at $M = 300$ and $\lambda = \SI{640}{nm}$, arranged according to the detector position.
    Panel b), the corresponding dataset of the autocorrelated PSFs.
    In panel c), we report the contributions to the $H_\text{SUM}$ separated into the autocorrelation components, $H^{AC}_\text{SUM}$, and into all the possible cross-correlation terms $H^{CC}_\text{SUM}$. Respect to $H^{AC}_\text{SUM}$, the $H^{CC}_\text{SUM}$ has always a wider support. The circle draws the full-width half maximum of the curve.
    Panel d) reports the contribution of the cross-correlation term of the PSFs aligned with APR, $H^{CC}_\text{APR}$. We may notice that realigning PSFs has the effect of making  $H^{CC}_\text{APR}$ more compact than $H^{CC}_\text{SUM}$.
    In e), our strategy instead neglects any cross-correlation components, aiming at forming a reconstruction only based on the average autocorrelation of the PSFs.
    Panel f) shows the line profile of the peak-normalized autocorrelated PSFs for the three considered cases.
    In panel g), we report the corresponding MTFs.
    }
    \label{fig:PSFscheme}
\end{figure}

Calculating and compensating for the shifts in the images of the ISM dataset might be challenging in those scenarios where the assumptions of the APR method cease to be valid, such as in the low magnification case.
In Fig. \ref{fig:PSFscheme}a, we report the simulated PSF map of a typical ISM system, where we may appreciate how the shape of each $h(\xs | \xd)$ change depending on the detector position with respect the optical axis.
To tackle this challenge, we need to look at the ISM reconstruction problem from a new perspective. Thus, we approach the image formation equation in the shift-space, by calculating the autocorrelation of the image. To simplify the notation, from now on we drop the explicit dependency from the $\xs$ and coordinate, and we substitute the $\xd$ variable with a discrete index. The auto-correlated image is
\begin{align}
    I_{\gamma} &= i_{\gamma} \star i_{\gamma} = \left(o \star o \right) * \left(h_{\gamma} \star h_{\gamma} \right)
\end{align}
where the capital letter indicate the autocorrelation of the corresponding quantity, $\star$ is the cross-correlation operator with respect to the coordinate $\xs$ and $\gamma\in\qty{\text{SUM, ISM}}$. Therefore, the autocorrelation of the sample $o$ is blurred by $H_\gamma$ (representative example shown in Fig. \ref{fig:PSFscheme}b), the autocorrelation of the real-space PSF (APSF).
For any of the above PSFs, the resulting autocorrelation includes all the terms
\begin{align}
    H &= \left( \sum_{\xd} h(\xs | \xd) \right) \star \left( \sum_{\vect x'_{d}} h(\xs | \vect x'_{d}) \right) = \sum_{\xd, \vect x'_{d}} h(\xs | \xd) \star h(\xs | \vect x'_{d}) = \nonumber \\
    &= \sum_{\xd} h(\xs | \xd) \star h(\xs | \xd) + \sum_{\xd \neq \vect x'_{d}} h(\xs | \xd) \star  h(\xs | \vect x'_{d}) = \nonumber \\
    &= H^{AC}(\xs) + H^{CC}(\xs),
    \label{eq:autopluscorr}
\end{align}
The two terms of Eq. \ref{eq:autopluscorr} represent the auto-correlation (AC) and cross-correlation (CC) components, respectively.
Using the shift-space representation, we can better understand why APR improves resolution, compared to the direct sum.
The $H^{CC}$ term is sensitive to differences in the PSFs -- such as the shift -- and its support becomes larger with increasingly large shifts between the PSFs.
Instead, the $H^{AC}$ term does not depend on shifts in real space because it includes only contributions from autocorrelations which are shift-invariant.
In the case of APR, the shifts between the PSFs are compensated in post-processing, minimizing the cross-correlation term $H^{AC}$ and preserving a small support for $H_\gamma$.
Since the Fourier dual of $H$ is the modulation transfer function (MTF), a small support of $H$ implies a large support of the MTF and a higher resolution of the image.
This situation can be easily visualized in Fig. \ref{fig:PSFscheme}c-d, where we report the peak normalized PSF components for SUM and APR, the latter having a more compact $H^{CC}$ term.

As anticipated, the PSFs may differ not only in position, but also in shape (Fig. \ref{fig:PSFscheme}a), depending on the experimental conditions.
Due to its algorithm design, APR already provides the most compact cross-correlation representation as a result of the phase-correlation process used to realign the images (see Eq. \ref{eq:phasecorr}).
Instead, working in the autocorrelation space, grants access to direct calculation of $H^{AC}$, and we can exclude the cross-correlation term from the sum (Fig. \ref{fig:PSFscheme}e).
The comparative profile analysis of the APSFs is presented in Fig. \ref{fig:PSFscheme}f, where we compare the ideal profiles of the pinhole detection (CLSM 0.4), SUM (CLSM 2.0), APR and ACO.
Among those, ACO reaches almost the resolution of the fully closed pinhole, still retaining the information coming from the entire array detection.
The profile of the MTFs (Fig. \ref{fig:PSFscheme}g) provide an even clearer picture on how ACO retains more frequencies compared to the other methods.
More precisely, we can build the autocorrelated image as:
\begin{align}
    I_{\text{ACO}} (\xs) &\overset{\operatorname{def}}{=} \sum_{\xd} i(\xs | \xd) \star i(\xs | \xd) = \nonumber \\
    &= \left[ o(\xs) \star o(\xs) \right] * \sum_{\xd} h(\xs | \xd) \star h(\xs | \xd) = O(\xs) * H_\text{ACO}(\xs)
\end{align}
Namely, we can build the sharpest autocorrelation by performing the summation of the ISM dataset in the autocorrelation space.
However, to obtain the final super-resolution image $i_{\text{ACO}}$, we need to invert this averaged $I_{\text{ACO}}$, which underlines an ill-posed problem:
\begin{equation}
    i_{\text{ACO}} = {\argmin_{i}} \, D\left( i \star i \, \| \, I_{\text{ACO}}\right)
    \label{eq:inverseproblem}
\end{equation}
where $D(\cdot\|\cdot)$ is a discrepancy metric which is ideally minimized when $I_{\text{ACO}} = i_{\text{ACO}} \star i_{\text{ACO}}$.
As we will detail in the following section, the numerical inversion of the autocorrelation requires care to be performed effectively. In ISM, to our advantage,  the inversion can be carried out reliably since we have a collection of direct observations of the sample suitable as a starting point for the inversion.

\subsection{Numerical inversion}
Having access to a better-resolved autocorrelation does not directly imply the capability of obtaining a sharper result in real space.
Although computing autocorrelation of real images is a simple forward operation, returning from the autocorrelation space to the image plane requires approaching a dedicated inverse problem (defined as in Eq. \ref{eq:inverseproblem}).
Since the autocorrelation is directly linked to the Fourier modulus via the Fourier spectrum theorem, the problem of inverting the autocorrelation falls into the class of the Fourier-phase retrieval problems \cite{shechtman2015phase}.
One of the most common strategies to solve phase retrieval problems is approaching the problem iteratively (such as in error reductions, hybrid input-output, and relative variations \cite{fienup1978reconstruction}), in which we alternate back and forth between real and Fourier space applying constraints to the recovered images.
Those methods are simple to implement and numerically efficient, but they suffer from image stagnation, are sensitive to noise, and could lead to twin-image artifacts \cite{fienup2013phase}.
A much more solid approach, the Schultz-Snyder fixed point iteration \cite{schulz1992image}, relies on the optimization of a generalization of the Kullkback-Leibler divergence, namely the Csiszár’s \textit{I}-divergence \cite{Csiszar1991}.
Given an autocorrelation signal, $I_\text{ACO}$, we may reconstruct the underlying imaged object $i_\text{ACO}$ via an iterative procedure:
\begin{equation}
    i_{\text{ACO}}^{\,t+1} = i_{\text{ACO}}^{\,t} \left[ i_{\text{ACO}}^{\,t}*\left(\frac{I_{\text{ACO}}}{i_{\text{ACO}}^{\,t} \star i_{\text{ACO}}^{\,t}}\right) + i_{\text{ACO}}^{\,t} \star \left(\frac{I_{\text{ACO}}}{i_{\text{ACO}}^{\,t} \star i_{\text{ACO}}^{\,t}}\right) \right],
    \label{eq:SS}
\end{equation}
where $t$ is the iteration index.
Schultz-Snyder's work has proposed this iterative approach \cite{schulz1992image} as a result of the optimization of the \textit{I}-divergence.
This method exhibits very stable numerical behavior, can incorporate support constraints, and preserves total intensities; however, it suffers from a slow convergence rate.
Furthermore, due to the non-convex nature of the phase retrieval problem, the final reconstruction strongly depends upon the algorithm initialization $i_\text{ACO}^{\,t=0}$ \cite{choi2006convergence}.
The above considerations have often restricted its usage in practical problems.
Nowadays, thanks to the usage of GPUs, approaching an autocorrelation inversion with the Schultz-Snyder becomes an interesting option that is worth investigating.
Due to its limited possibilities to explore the loss-landscape in search of the optimal minimum, we will use this method as a refiner, as we can easily compute an optimal starting point for the algorithm.
Differently from crystallography or astronomical imaging -- where we have no direct way to observe a low-resolution version of the object -- here we calculate the autocorrelation starting from a direct measurement of the object of interest.
In principle, any combination of these images could be used as a starting point for the optimization because it provides a low-resolution version of the sample, alleviating the burden of exploring the $I$-divergence landscape in search of the global minimum.
Thus, in the context of ISM, the APR image is the best candidate for initializing the autocorrelation inversion thanks to higher resolution and SNR, and consequently, we start the iterations with $i_\text{ACO}^{\, t=0} = i_\text{APR}$.
The proposed pipeline can be summarized as follows. We leverage the insights reported in Sec. \ref{sec:theory} to compute the autocorrelation of each image in the map and sum them.
Therefore, we obtain the sharpest autocorrelation signal using all the information coming from the detector array (Fig. \ref{fig:opticalsetup}b).
Finally, we reconstruct the final image using the inversion algorithm described by Eq. \ref{eq:SS} using the APR reconstruction as the starting point, minimizing the issues related to the non-convexity of the algorithm.
Given these initial conditions, we consistently obtained an extremely stable behavior, and the algorithm did not produce any detectable artifact even after an extremely long run of iterations ($10^7$).

\section{Results}

\subsection{Performance analysis at different simulated magnifications}
\begin{figure}
    \centering
    \includegraphics[width=1.\textwidth]{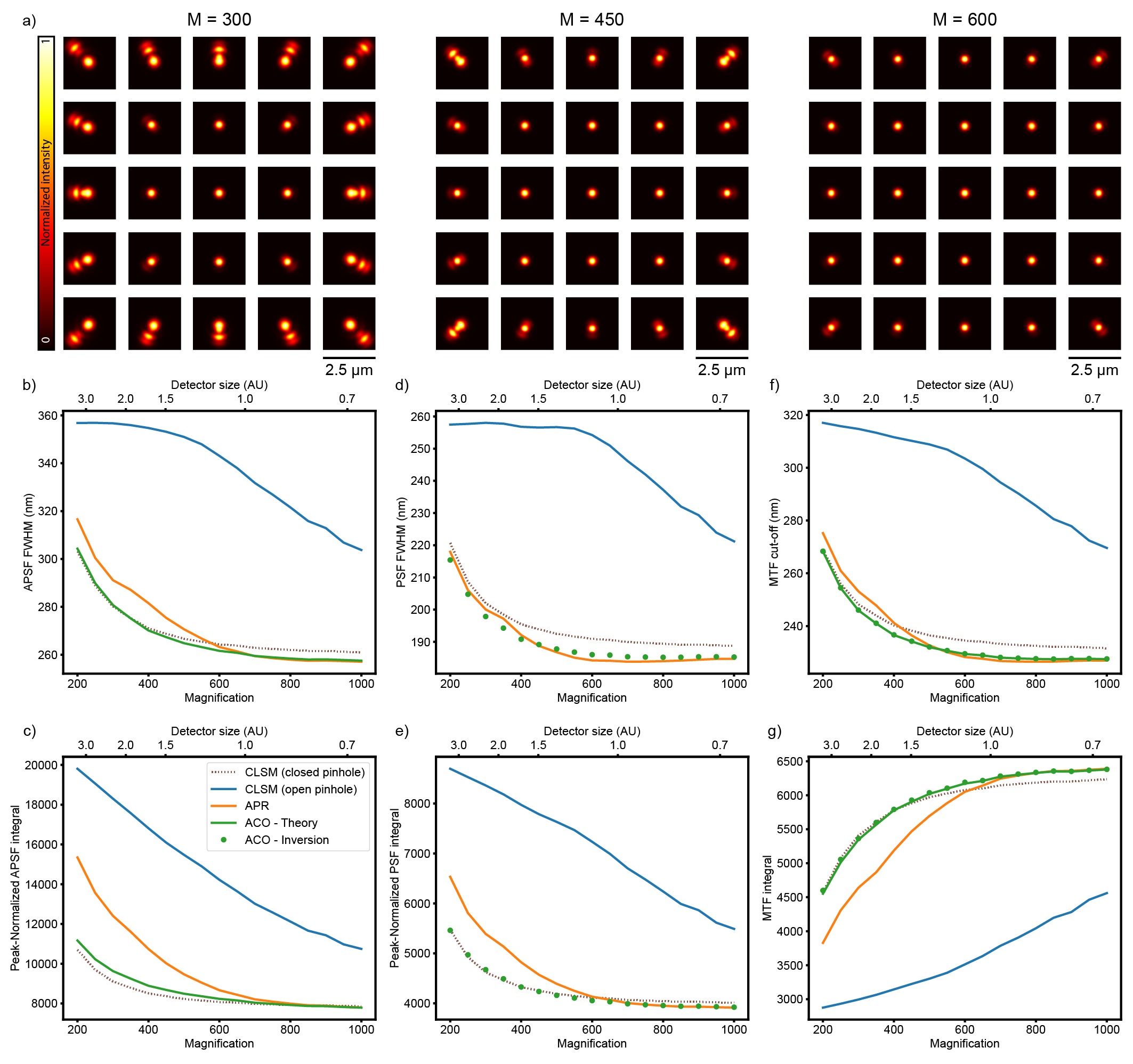}
    \caption{\textbf{Reconstruction performance at various magnifications.}
    PSF datasets simulated at $\lambda = \SI{640}{nm}$ and magnification of 300, 450, and 600. The larger the magnification, the more structurally similar the PSFs (a).
    The resolution provided by different reconstruction techniques is estimated at different magnifications as the FWHM of the autocorrelated PSFs (b), as the FWHM of the reconstructed PSF (d), and as the cut-off of the MTF, calculated using a threshold of 10\%.
    We also computed the integral of the autocorrelated PSF (c), the reconstructed PSF (e), and the MTF (g). The maximum of each integrand is normalized to unity.
    We estimated the performance of ACO-ISM directly from the autocorrelation space (green continuous line) or after the reconstruction of the PSF in real space (green dots). The size of the closed pinhole is one-fifth of the size of the detector array.
    }
    \label{fig:magnificationstudy}
\end{figure}

The PSFs of an image-scanning microscope sensibly depend upon the magnification with which we observe the object of interest and some hardware features of the detector array.
For this reason, we begin our study by assessing the resolution gain provided by our method in non-noisy scenarios at various magnification levels.
To this end, we simulate the PSFs of an ideal $5\times 5$ array detector at various magnification factors using the BrightEyes-ISM software \cite{Zunino2023b}.
Three representative examples of the simulated PSF dataset are presented in Fig. \ref{fig:magnificationstudy}a, where we can appreciate how, at high magnification, the PSF shape varies very little compared to that of the central detector.
First, we sum these PSFs to form the $h_\text{SUM}$, the PSF of the fully opened pinhole as large as the size of the detector array.
Subsequently, we apply APR to each PSF dataset, correcting the reciprocal shifts between the $h(\xs | \xd)$ and forming the $h_\text{APR}$.
Last, we compute the autocorrelation dataset that we use to obtain $H_\text{ACO}$.
We invert this latter with the Schultz-Snyder method (Eq. \ref{eq:SS}), obtaining the $h_\text{ACO}$ as a function of the varying magnification.

We begin our comparison by examining the APSFs at varying magnification, estimating the full-width at half-maximum (FWHM) of the angularly-integrated curve and the peak-normalized APSF integral (Fig. \ref{fig:magnificationstudy}b-c).
As expected, the ACO consistently exhibits the sharpest APSF and converges to APR at high magnifications when all the PSFs approach the same shape (above $600 \times$).
Similarly, we study the FWHM and the peak-integral of the PSFs for a more standard comparison (Fig. \ref{fig:magnificationstudy}d-e).
Regarding the FWHM of the PSF, APR and ACO exhibit a similar trend.
The integral, instead, gives a more comprehensive resolution metric, taking into account the contribution of the entire PSF shape.
Here, similarly to the APSF case, ACO always produces the sharpest PSF, converging to the APR result at high magnifications.
Additionally, to further characterize the resolution enhancement of these techniques, we estimate the modulation transfer function (MTF) of the simulated optical system.
Since the MTF can be directly calculated from the APR, we use the formulation on the right side of Eq. \ref{eq:MTF} to ensure that the results from the numerical inversions are consistent with the direct calculation and that no artefacts are introduced during the inversion process.
With the MFTs at hand, we compute the $10\%$ cut-off of the radially averaged profile and the MTF integral as a function of the magnification, which we show in Fig. \ref{fig:magnificationstudy}f-g.
The MTF cut-off indicates that ACO provides imaging gain at low magnification, converging to APR at high magnification.
Then, we analyze the global frequency contribution the imaging processing allows by calculating the peak-normalized MTF integral.
In this case, we verify that ACO robustly enables more frequencies than SUM and APR up to around $700\times$.
In these curves, we also include the trend from the central element of the $5\times 5$ array alone (dashed line), which produces an image similar to a confocal microscope with a small pinhole.
Consistently with the results reported in the literature \cite{Cox1982, Sheppard2013, Roth:16}, ISM achieves performance better than a fully-closed pinhole both with the APR and ACO reconstructions.
It is also worth mentioning that in the magnification range $500-700\times$, ACO globally allows more frequencies than either APR or CLSM with the closed pinhole.

All these results help us delineate the regime at which the autocorrelation inversion will predominantly boost resolution.
Low magnification implies significant PSF variability in shape, and in this situation, ACO could offer a great tool to improve imaging details only by performing post-processing operations.
At high magnification (thus, less PSF diversity), the resolution of the ACO method converges to APR because the cross-correlation component approaches a structure similar to the autocorrelation of the pinhole PSF.
However, these are ideal measurement scenarios in which the absence of noise lets us draw only theoretical predictions about pure resolution.
In a more realistic case, the noise will play a critical role, ultimately impacting the effective imaging resolution achieved by either APR or ACO. In the following section, we will test our method under experimental conditions to provide concrete examples of the benefits of its usage also with the presence of real noise.

\subsection{Experimental performance analysis varying detection wavelength}
\label{sec:expPSF}
In the previous section, we delved into how changing the magnification could influence the shape of the point spread functions (PSFs), thereby affecting the resolution of the techniques analyzed. However, in multi-colour imaging experiments, we typically rely on the detection of different wavelengths using a single detection objective and fixed magnification.
In this scenario, the wavelength variation alters the effective size of the detector array, as when changing the magnification. Thus, the modification of the shape and size of the PSFs may impact the reconstruction performance.

To assess the performance enhancement of ACO-ISM, we captured experimental images of gold beads with different excitation lasers ($\lambda =\SI{488}{nm}$, $\lambda =\SI{561}{nm}$, and $\lambda =\SI{640}{nm}$) at $M=450\times$, and carried out a resolution study similar to the previous case.
First, we sum the dataset to obtain an experimental estimation of the diffraction-limited PSF $h_\text{SUM}$.
Consequently, we process the dataset with the APR to obtain the PSF under experimental conditions, $h_\text{APR}$.
Lastly, we autocorrelate the $5\times 5$ dataset and average them, obtaining $H_\text{ACO}$.
As for the simulation study, such APSF is inverted using Eq. \ref{eq:SS} starting with the $h_\text{APR}$, obtaining the corresponding experimental $h_\text{ACO}$ of the imaging system processed using ACO-ISM.

\begin{figure}
    \centering
    \includegraphics[width=1.\textwidth]{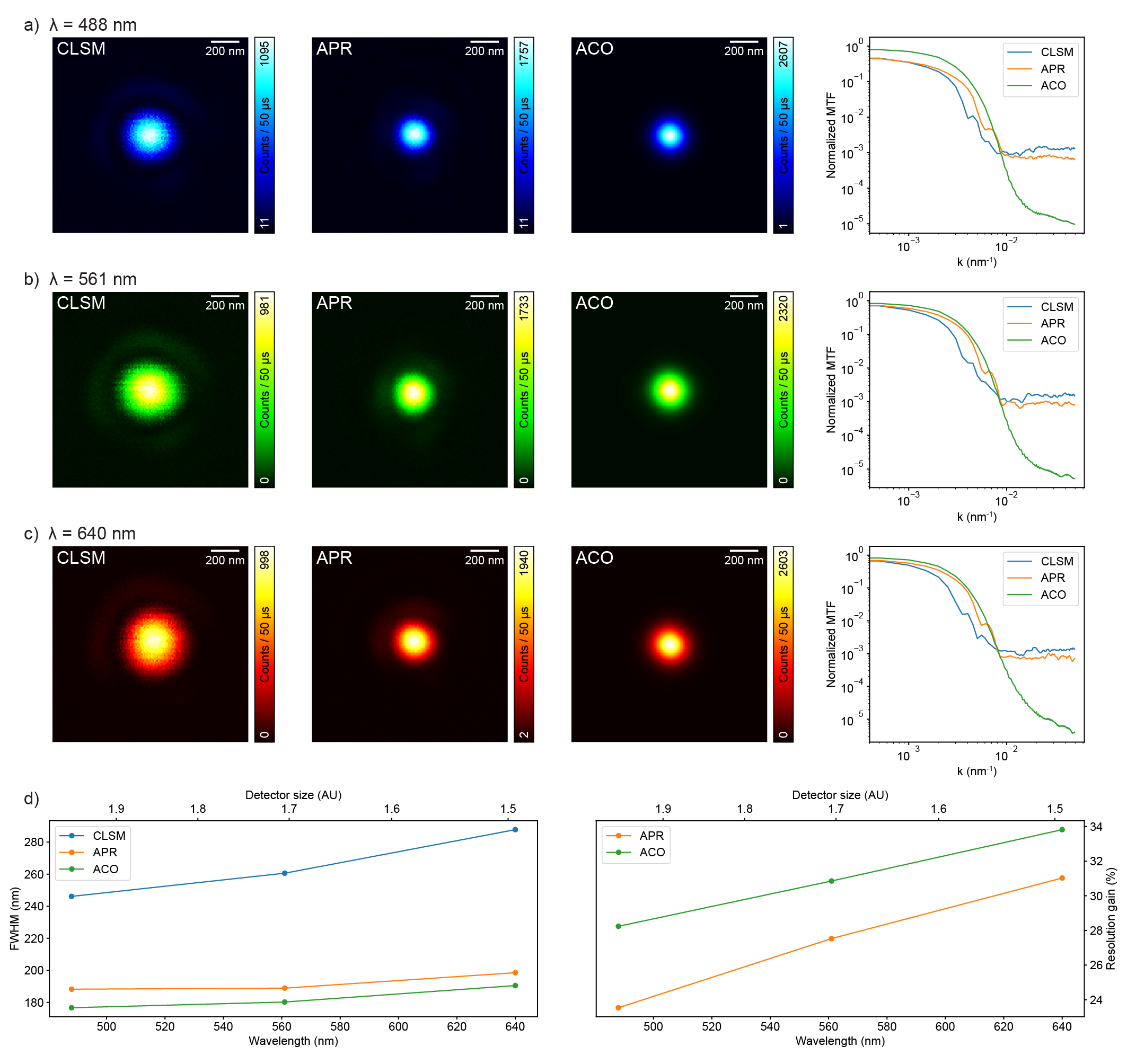}
    \caption{\textbf{Reconstruction of experimental point spread functions.} Reconstructed images of a gold bead ($\diameter = \SI{80}{nm}$) and the corresponding MTFs, acquired at the laser wavelength of $\lambda =\SI{488}{nm}$ (a), $\lambda =\SI{561}{nm}$ (b), and $\lambda =\SI{640}{nm}$ (c). Panel (d) shows the absolute (left) and relative (right) resolution (FWHM) of the reconstruction methods as a function of the wavelength or -- equivalently -- of the effective size of the detector array. The resolution gain is calculated relative to the open pinhole confocal measurement.}
    \label{fig:experimental_psfs}
\end{figure}

We begin our analysis with the shortest wavelength $\lambda=\SI{488}{nm}$, presenting the PSFs obtained in Fig. \ref{fig:experimental_psfs}a, alongside their corresponding MTF.
Even under experimental conditions, ACO gives access to a sharp PSF that extends the range of frequencies in the MTF than the other two methods.
Additionally, we notice a strong denoising effect which appears as an attenuation of the high-frequency spectrum beyond the support of the MTF.
This behaviour is intrinsic to the autocorrelation analysis since additive uncorrelated noise mainly contributes to the zero-shift pixel of the autocorrelation image.
A similar situation consistently happens for the other wavelengths studied $\lambda=\SI{561}{nm}$ (Fig. \ref{fig:experimental_psfs}b) and $\lambda=\SI{640}{nm}$ (Fig. \ref{fig:experimental_psfs}c), as it can also be seen by the fact that high-frequency components of the MTFs are efficiently attenuated.
As a result, the FWHM of the $h_\text{ACO}$ is always lower than that of APR, providing a constant resolution gain across all the wavelengths studied.
Furthermore, the secondary diffraction lobes surrounding the Airy disk of the APR and SUM reconstructions get effectively suppressed with the ACO analysis. Such an effect is attributable to the exclusion of cross-correlation components between PSF with differently oriented lobes.

Based on the findings of this study, it is evident that ACO-ISM enhances resolution gain and effectively reduces noise in processed images.
However, it is important to note that beads are simple objects with symmetrical structures having autocorrelations that are structurally alike.
In this case, the autocorrelation inversion is a straightforward problem compared to a more general case of structurally complex objects.
Therefore, in the following section, we evaluate the efficacy of our reconstruction framework in a more realistic imaging scenario.

\subsection{Experimental image reconstruction}
\begin{figure}
    \centering
    \includegraphics[width=\linewidth]{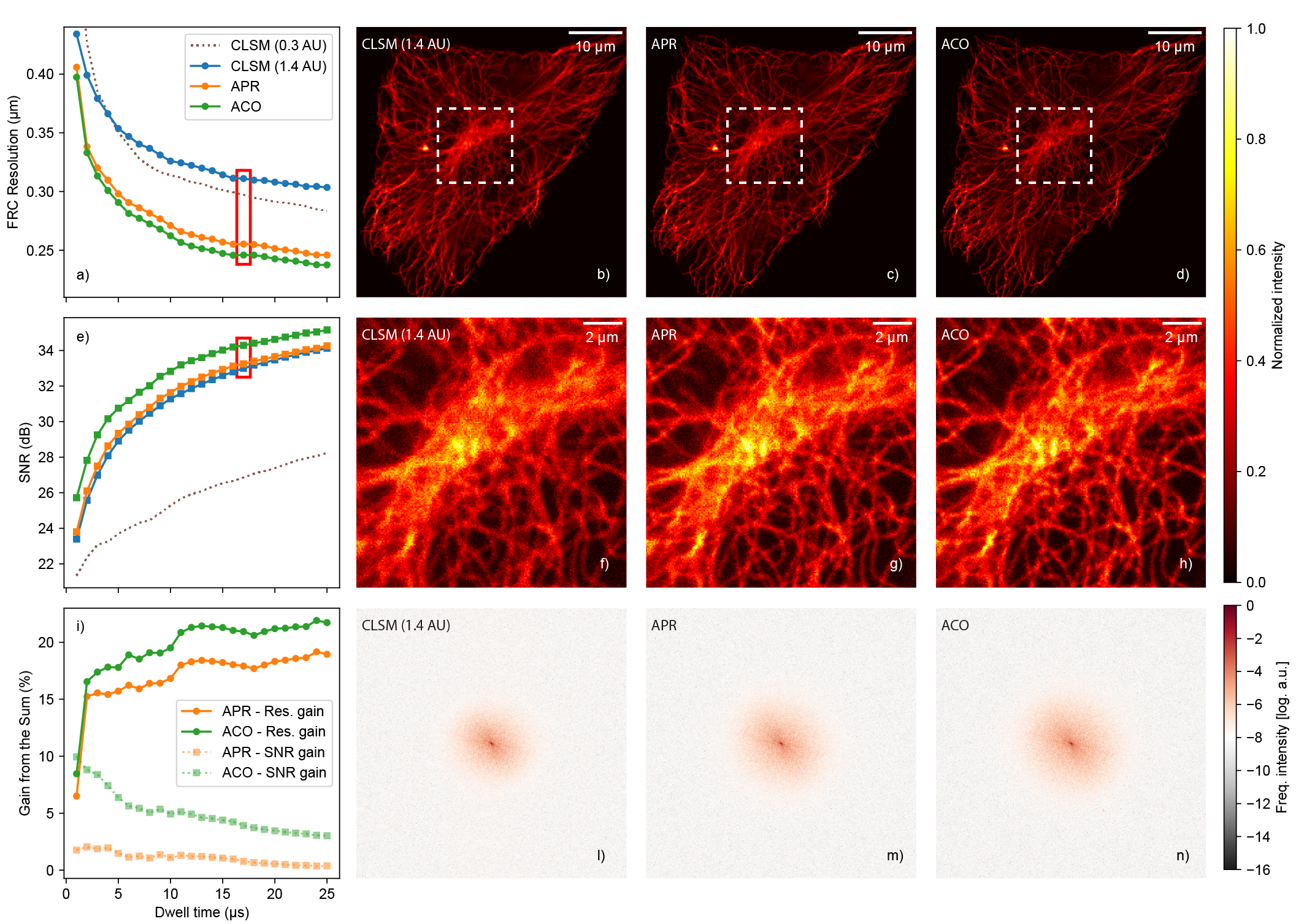}
    \caption{\textbf{ACO-ISM reconstruction comparison}
    Panel a), resolution estimated with FRC analysis varying the dwell time for the three methods.
    We notice the improved ACO curve against standard procedures.
    Visualization of the reconstruction result using the Sum (panel b,f), APR (panel c,g) and ACO inversion (panel d,h) and their corresponding zoomed visualizations.
    Panel i), study of the SNR as a function of the dwell time. Since APR does not account for noise removal, the gain achieved in terms of SNR is modest. The autocorrelation inversion instead also suppresses noise, sensibly improving SNR.
    Panel i), resolution (dots) and SNR (transparent squares) gains for the two methods (APR, ACO) with respect to the result obtained with standard SUM.
    Panel l), Fourier modulus of the $f_{Sum}$, m) Fourier modulus of the $f_{APR}$ and n) of the $f_{ACO}$.
    We notice how the frequency support expands, determining the resolution improvement of ACO reconstructions.
    }
    \label{fig:scheme}
\end{figure}

To validate the robustness of our approach, we run independent image reconstructions at increasingly higher SNR as a function of the dwell time.
For this imaging experiment, we used the same magnification as before, ($M=450\times$) and the excitation wavelength of $\lambda=640nm$.
Thanks to the high speed of the SPAD array detector, we acquired fifty ISM datasets of the same sample using a pixel dwell time \SI{1}{\micro s}. Cumulative summing the repetitions would give us a dataset with an equivalent pixel dwell time ranging from 1 to \SI{50}{\micro s}.
To estimate the resolution of those images, we make use of the Fourier ring correlation (FRC) method, which requires two independent realizations of the same images to estimate the frequency cutoff \cite{Nieuwenhuizen2013, Tortarolo2018, Koho2019}. For this reason, we split the datasets into two separate collections, obtaining two independent images per dwell time, up to \SI{25}{\micro s}.
First, we consider the sum of all these images without further data manipulation, producing an image with the same resolution as it would be detected with a fully opened confocal pinhole (Fig. \ref{fig:scheme}b,f).
The FRC curve is presented with the blue line in Fig. \ref{fig:scheme}a, where we observe a significant increase in the image quality at a longer exposure time, as expected.
Indeed, the FRC algorithm calculates the resolution as the inverse of the highest spatial frequency above the noise level.
Instead of summing all the images in the $5 \times 5$ dataset, APR recovers the optimal alignment between them so that their sum produces a substantially better image (Fig. \ref{fig:scheme}c,g).
FRC captures the enriched details of APR reconstructed images by returning an improved resolution curve (orange plot) compared to the sum.
Starting from the APR result and optimizing the $I$-divergence further enriches the details of the reconstructed image (Fig. \ref{fig:scheme}d,h).
We observe a gain in resolution at any dwell time investigated, demonstrating a robust benefit with a simple post-processing strategy without any experimental characterization of the PSF.
In fact, the autocorrelation model implicitly underlines a sharper PSF without the need to measure it.

In our analysis, it is also worth evaluating the amount of noise in the images.
To do so, we decided to measure the signal-to-noise ratio (SNR) defined in Eq. \ref{eq:SNR} as a function of the dwell time (Fig. \ref{fig:scheme}e).
As expected, the SNR of the fully opened array increases the longer we expose the imaged object.
APR has a limited influence on the SNR value since we are only summing the images at their optimal cross-correlation position, increasing the signal contrast versus the noise level.
With ACO instead, we have an active denoising effect that lifts the signal coming from the object, suppressing at the same time the background noise (green curve in Fig. \ref{fig:scheme}e).
This behaviour confirms the experimental findings obtained during the PSF characterization.
The autocorrelations are, in fact, remarkably insensitive to uncorrelated noise since the autocorrelation of random noise is a sharp peak located in the centre \cite{ancora2020deconvolved}.
At the expense of carrying out a single inversion task, we are, thus, ameliorating both the resolution and SNR of the finally reconstructed image.
The relative gain for each quantity of interest is reported in Fig. \ref{fig:scheme}i against the sum image, and it is consistent along the whole noise range studied.
To further consider the structural effect of the ACO processing on the final images, we report the Fourier modulus of each representative image (Fig. \ref{fig:scheme}l-n), where it is possible to appreciate how the information content expands in $i_\text{ACO}$, enriching the content of high-frequency components.

As predicted from the PSF studies, ACO-ISM then enables the formation of a sharper image by simply processing the same dataset in a smart way without specific characterization of the setup.
It does so by simultaneously acting on the resolution and on the noise level, returning naturally looking images that do not exhibit artefacts.
This makes our technique very robust, permitting flexibility to reprocess previously acquired data for further augmentation of the resolution.

\section{Conclusions}
In this work, we presented a novel reconstruction framework that exploits the properties of the autocorrelation operator to form high-resolution and low-noise images in ISM measurements.
We begin our study by analyzing the autocorrelation properties of simulated PSFs to establish a solid theoretical basis for the approach.
By doing so, we pointed out that the simple fact of realigning images acts by making the support of the cross-correlation component more compact than that of the direct sum: this operation justifies the resolution increase of the pixel reassignment technique.
Instead, when dealing with PSFs that are changing across the array detectors, the cross-correlation term cannot be made as compact as the APSF, determining a practical deviation of the resolution achieved from theoretical predictions.
This resolution loss consistently happens in many experimental conditions, depending upon magnification and imaging wavelength in a nontrivial way.
To beat this current limitation in image processing, we calculated the autocorrelation component from the image maps by averaging in the shift space, explicitly neglecting cross-correlation components.
Indeed, the MTF analysis revealed that this operation allows for more frequencies to form the image, only requiring us to get back to the real space to finalize its reconstruction.
To validate the theoretical predictions, we tested our algorithm on simulated and real ISM datasets, confirming the benefits of our proposed post-processing strategy and demonstrating significant improvements in terms of resolution and SNR.
The effectiveness of autocorrelation inversion was also tested across various pixel dwell times, showcasing its reliability under various imaging scenarios.

Yet, several improvements can still be made, particularly regarding how we approach the autocorrelation inversion.
Indeed, the starting point and the convergence rate for the inversion algorithm are sensitive matters \cite{choi2006convergence}.
For example, neural networks exhibit promising performances in phase retrieval problems \cite{pellegrini2023phase} and may solve some issues connected to slow convergence and optimal exploration of the reconstruction landscape.
Resolution increase could also be achieved by incorporating informed \cite{ancora2020deconvolved} or blind \cite{corbetta2022blind} deconvolution in the autocorrelation inversion to extend imaging potentials further.
As seen in Sec. \ref{sec:expPSF}, inverting autocorrelated PSFs provides a much smoother version of the PSF, effectively denoising it. Thus, the inversion of the autocorrelation could also be useful in downstream analysis, especially if followed by (or incorporating as in \cite{ancora2020deconvolved}) a deconvolution step.
Despite focusing our analysis on the case of conventional ISM, any laser scanning technique that exploits a detector array --- such as non-linear ISM \cite{Koho2020}, STED-ISM \cite{Tortarolo2022}, SOFISM \cite{Sroda2020} or quantum ISM \cite{Tenne2019} -- can benefit from our innovative framework.
The ACO-ISM approach also has the potential to effectively deal with sample-dependent aberrations. These aberrations are caused by imaging depth, sample position, and local refractive properties, which may further introduce additional PSF variation among the detector elements. Our proposed framework is specifically designed to tackle these challenges, and we are planning to investigate further in this direction.
Additionally, exploring the feasibility of our approach also extends to architectures different from ISM. Any imaging system relying on the acquisition of multiple images could benefit from this pipeline, for example, in STED with 1D depletion patterns \cite{Kruger:20, Kratz:22} or multi-view confocal microscopy \cite{wu2021multiview}, to cite a few.
We believe, then, that this novel approach for image reconstruction has the potential to reliably enter many imaging pipelines, paving the way for potential advancements in biomedical imaging and beyond.

\section{Materials and methods}

\subsection*{Microscope architecture}
For this work, we used a custom-built ISM setup. The excitation beams are provided by three triggerable pulsed (\SI{80}{ps} pulse-width) diode lasers emitting at \SI{640}{nm}, \SI{561}{nm}, and \SI{488}{nm} (LDH-D-C-640, LDH-D-C-560, and LDH-D-C-488 -- Picoquant). We finely control the power of each laser source using an acoustic optical modulator (AOM, MT80-A1-VIS, AAopto-electronic).
Each laser beam is individually coupled into a single-mode polarisation maintaining fibre (PMF) to enhance the spatial quality of the beam. We use a half-wave plate (HWP) to align the polarization axis of the beam to the fast axis of the PMF.
A set of dichroic mirrors (491 short-pass, 590 short-pass, 750 short-pass) allows for the co-alignment of the three laser beams. A multi-band dichroic mirror (488-560-640-775) separates the excitation and fluorescence light.
Two galvanometer scanning mirrors (6215HM40B, CT Cambridge Technology), a scan lens and a tube lens -- of a commercial confocal microscope (C2, Nikon) -- deflect and direct all the beam towards the objective lens (CFI Plan Apo VC 60$\times$, 1.4 NA, Oil, Nikon) to perform the raster scan on the specimen.
The objective lens is mounted over a nanopositioner (FOC.100, Piezoconcept), enabling z-scanning.
The fluorescence light is collected by the same objective lens, de-scanned, and sent towards the detection path. This latter consists of a set of lenses to form a telescopic system that conjugates the sample plane onto the detector plane with an overall magnification of 450$\times$.
Fluorescence light is selected by a dedicated set of notch and bandpass filters, depending on the excitation wavelength (red set: ZET633TopNotch and ET685/70M). No filters are used for imaging scattering samples.
The detector is a $5 \times 5$ SPAD array (PRISM-light kit, TTL version, Genoa Instruments) with a pixel pitch of \SI{75}{\micro m}. The TTL signal generated by each SPAD element is read by a multifunction FPGA-based I/O device (NI USB-7856R from National Instruments), which acts both as a DAQ system and a control unit. The BrightEyes-MCS software \cite{mcs} controls the entire microscope, including the galvanometric mirrors, the FOC, and the AOMs. The software also provides real-time visualization of the image during the scan and saves the raw data in a hierarchical data format (HDF5) file.

\subsection*{Sample preparation}

Human HeLa cells were fixed with ice methanol, 20 minutes at \SI{-20}{\celsius} and then washed three times for 15 minutes in PBS. After 1 hour at room temperature, the cells were treated in a solution of 3\% bovine serum albumin (BSA) and 0.1\% Triton in PBS (blocking buffer). The cells were then incubated with the monoclonal mouse anti-$\alpha$-tubulin antiserum (Sigma Aldrich) diluted in a blocking buffer (1:800) for 1 hour at room temperature. The $\alpha$-tubulin antibody was revealed using Abberior STAR Red goat anti-mouse (Abberior). The cells were rinsed three times in PBS for 5 minutes.

\subsection*{Numerical simulations}

\noindent We simulated the PSFs of the ISM system using the BrightEyes-ISM Python package \cite{Zunino2023b}. The computation is performed using a vectorial diffraction model \cite{Caprile2022}. We simulated the detector array as a set of 25 squared pinholes arranged in a square matrix, using a pixel size of \SI{75}{\micro m} and neglecting the dead area around each detector. The excitation and emission wavelengths are, respectively, $\lambda_{\exc} = \SI{640}{nm}$ and $\lambda_{\emi} = \SI{660}{nm}$. We set the numerical aperture of the objective lens to $\NA = 1.4$ and the refractive index of the immersion oil to $n = 1.5$. The magnification value may vary from 200 to 1000.

\subsection*{Image processing}

\paragraph{Fourier Ring Correlation.} Given two independent realizations of the same image $i_1$ and $i_2$, we compute the corresponding Fourier transformed images $\tilde{\imath}_1$ and $\tilde{\imath}_2$. Then, we calculate the FRC as the following normalized cross-correlation function
\begin{equation}
    \mathrm{FRC}\left(q\right)=\frac{\sum\limits_{k_{x}, k_{y} \in q}
    \tilde{\imath}_1\left(k_{x}, k_{y}\right) \overline{\tilde{\imath}_2\left(k_{x}, k_{y}\right)}}
    {\sqrt{\sum\limits_{k_{x}, k_{y} \in q}
    \left|\tilde{\imath}_1\left(k_{x}, k_{y}\right)\right|^{2} \sum\limits_{k_{x}, k_{y} \in q}
    \left|\tilde{\imath}_2\left(k_{x}, k_{y}\right)\right|^{2}}}
\end{equation}
where the overline represents the complex conjugate, $k_x$ and $k_y$ are, respectively, the horizontal and vertical spatial frequencies, and $q=\sqrt{k^2_x+k^2_y}$ is the radial spatial frequency. We fit the resulting curves to the following sigmoid model
\begin{equation}
    \sigma(q) = \frac{1}{ 1 + \exp\qty(\frac{q-t}{s}) } + b
\end{equation}
where $t$, $s$, and $b$ are fitting parameters. In detail, $b$ describes a possible offset of the FRC curve due to partially correlated high-frequency noise, such as the mechanical noise of the scanners. We subtract the offset from the FRC curve and rescale the result such that it spans the $[0, 1]$ range. Finally, we calculate the effective resolution using the $1/7$ threshold criterion.

\paragraph{Calculation of confocal images.} The open pinhole CLSM image is obtained from the raw ISM dataset by summing all the images
\begin{equation}
    i_{\text{open}}(\xs) = \sum_{\xd} i(\xs | \xd)
\end{equation}
The closed pinhole CLSM image is obtained by selecting only the image collected by the central element of the detector array
\begin{equation}
    i_{\text{closed}}(\xs) = i(\xs | \vect 0)
\end{equation}

\paragraph{Adaptive pixel reassignment.} The APR algorithm calculates the shift vectors $\vect \mu(\vect x_d)$ and shifts back each scanned image of the corresponding quantity
\begin{equation}
    i(\vect x_s | \vect x_d) \xrightarrow[]{}  i\qty[\vect x_s + \vect \mu(\vect x_d) | \vect x_d]
\end{equation}
The final ISM image is calculated as

\begin{equation}
    i_{\text{ISM}}(\xs) = \sum_{\xd} i(\xs + \vect \mu(\vect x_d) | \xd)
\end{equation}
We calculated the shift vectors using the following phase correlation algorithm

\begin{equation}
    \vect \mu( \vect x_d ) = \argmax_{ \vect x_s}  \FT^{-1} \qty{ \frac{ \FT\qty{i(\vect x_s|\vect x_d)}   \cdot \overline{\FT\qty{i(\vect x_s|\vect 0)}} } { \qty| \FT\qty{i(\vect x_s|\vect x_d)} \cdot \overline{\FT\qty{i(\vect x_s|\vect 0)}} | } }
    \label{eq:phasecorr}
\end{equation}

\paragraph{Autocorrelation and autocorrelation inversion.}
When computing the autocorrelation, we first subtract the minimum value of the image and then we mask out the value of the zero-shift in the autocorrelation domain.
Additive uncorrelated noise predominately enters this region of the autocorrelation, so we replace the value of the central pixel with harmonic inpainting.
This techniques estimates the value of a masked region (in our case, the single pixel at the zero-shift position) based on surrounding values.
For this task, we use the python a specific function from the skimage library, the \texttt{restoration.inpaint\_biharmonic}.

After this pre-processing step, we invert the average autocorrelation $I_\text{ACO}$ using the Eq. \ref{eq:SS}, starting with the APR reconstruction $i_\text{ACO}^{\,t=0}=i_\text{APR}$ and iteratively refining it.
The number of iterations required for the convergence of the algorithm is data-dependent.
As a convergence criterion, we monitor the root mean square error between the autocorrelation of the reconstructed image and the averaged autocorrelation.
Experimentally we found that structured images typically converge after long runs, consisting of $10^5-10^7$ iterations.
Poorly structured images instead, such as simulated PSF data, converge substantially faster, requiring $10^3$ iterations.
With experimental PSF, the convergence was found after only $10-10^2$ iterations.

\paragraph{Signal to noise ratio.}
For the SNR, we use the definition:
\begin{equation}
    \text{SNR} = 10 \log_{10} \left(  \frac{ \left\langle i_\epsilon^2 \right\rangle }{ \left\langle \epsilon^2 \right\rangle} \right)
    \label{eq:SNR}
\end{equation}
where $i_\epsilon = i + \epsilon$ being $\epsilon$ the additive noise that perturbs the original image $i$ and $\left\langle \cdot \right\rangle$ the arithmetic mean operator.
We calculated the mean square of the noise over an empty region of the image, where no fluorescence signal is recorded.

\paragraph{Modulation Transfer Function.}
The modulation transfer function (MTF) can be computed either starting from the PSF, $h_\gamma$ with $\gamma=\left\{\text{SUM},\text{APR},\text{ACO} \right\}$ directly:
\begin{equation}
    \text{MTF}_\gamma = \left| \mathcal{F}\left[h_\gamma \right] \right|
\end{equation}
or also from the autocorrelation APSF, $H_\gamma$:
\begin{equation}
    \text{MTF}_\gamma = \sqrt{ \mathcal{F}\left[H_\gamma \right] }.
    \label{eq:MTF}
\end{equation}

\paragraph{FWHM estimation.} In order to estimate the FWHM of a PSF, we fitted the image to a multivariate normal distribution

\begin{equation}
    g(\boldsymbol{x}|\boldsymbol{\mu}, \boldsymbol{\Sigma}) = A\exp\left(-\frac 1 2 \left({\mathbf x} - {\boldsymbol\mu}\right)^\mathrm{T}{\boldsymbol\Sigma}^{-1}\left({\mathbf x}-{\boldsymbol\mu}\right)\right)
\end{equation}
where $A$ is a scalar coefficient and
\begin{equation}
    \boldsymbol\mu = \begin{pmatrix} \mu_x \\ \mu_y \end{pmatrix} \qquad
    \boldsymbol\Sigma = \begin{pmatrix} \sigma_x^2 & \rho \\
                             \rho  & \sigma_y^2 \end{pmatrix}.
\end{equation}
are the mean vector and covariance matrix, respectively.
We estimated the average FWHM as follows
\begin{equation}
    \text{FWHM} = 2 \sqrt{ \ln{2} \Tr \boldsymbol S }
\end{equation}
where $\boldsymbol S$ is the diagonalized covariance matrix.

\section*{Acknowledgment}
D.A. has received funding from the European Union’s Horizon 2020 research and innovation programme under the Marie Skłodowska-Curie grant agreement No. 945405 (ARISE). G.V. received funding from the European Research Council, \textit{BrightEyes}, ERC-CoG No. 818699.

\section*{Disclosures}
G.V. has a personal financial interest (co-founder) in Genoa Instruments, Italy.

\section*{Bibliography}
\bibliography{bibliography.bib}

\end{document}